\documentclass[twocolumn,epjc3]{svjour3}

\RequirePackage[T1]{fontenc}
\usepackage{orcidlink}

\smartqed  

\RequirePackage{graphicx}
\RequirePackage{mathptmx}      
\RequirePackage{flushend}
\RequirePackage[numbers,sort&compress]{natbib}

\hypersetup{
    colorlinks=true,
    citecolor=blue,
    urlcolor=blue,
    linkcolor=blue
}

\usepackage{amsmath} 
\usepackage{lineno} 

\journalname{Eur. Phys. J. C}

\begin{document}

\title{Study of two-photon femtoscopic correlations in Ag+Ag collisions at $\mathbf{\sqrt{s_{NN}}}$ = 2.55 GeV with HADES}

\author{HADES collaboration \\[5bp]
R.~Abou~Yassine$^{7,16}$, J.~Adamczewski-Musch\orcidlink{0000-0002-7586-8504}$^{6}$, C.~Asal$^{9}$, M.~Becker$^{12}$,
A.~Belounnas$^{16}$, A.~Blanco\orcidlink{0000-0001-9827-8294}$^{2}$, C.~Blume\orcidlink{0000-0002-6800-3465}$^{9,6,f}$, Y.~Bondar\orcidlink{0000-0003-2773-9668}$^{4}$, L.~Chlad\orcidlink{0000-0003-3814-2920}$^{17,h}$,
P.~Chudoba\orcidlink{0000-0002-7640-3418}$^{17,h}$
% ...
}

\institute{
\mbox{} \\[-8bp]
\mbox{$^{1}$Ruhr-Universit\"{a}t Bochum, 44801~Bochum, Germany}\\
\mbox{$^{2}$LIP-Laborat\'{o}rio de Instrumenta\c{c}\~{a}o e F\'{\i}sica Experimental de Part\'{\i}culas, 3004-516~Coimbra, Portugal}\\
\mbox{$^{3}$AGH University of Krakow, Faculty of Physics and Applied Computer Science, 30-059~Krakow, Poland}\\
\mbox{$^{4}$Institute of Nuclear Physics, Polish Academy of Sciences, 31342~Krak\'{o}w, Poland}\\
\mbox{$^{5}$Smoluchowski Institute of Physics, Jagiellonian University of Cracow, 30-059~Krak\'{o}w, Poland}\\
\mbox{$^{6}$GSI Helmholtzzentrum f\"{u}r Schwerionenforschung GmbH, 64291~Darmstadt, Germany}\\
\mbox{$^{7}$Institut f\"{u}r Kernphysik, Technische Universit\"{a}t Darmstadt, 64289~Darmstadt, Germany}\\
\mbox{$^{8}$Institut f\"{u}r Strahlenphysik, Helmholtz-Zentrum Dresden-Rossendorf, 01314~Dresden, Germany}\\
\mbox{$^{9}$Institut f\"{u}r Kernphysik, Goethe-Universit\"{a}t, 60438 ~Frankfurt, Germany}\\
\mbox{$^{10}$Excellence Cluster 'Origin and Structure of the Universe' , 85748~Garching, Germany}\\
\mbox{$^{11}$Physik Department E62, Technische Universit\"{a}t M\"{u}nchen, 85748~Garching, Germany}\\
\mbox{$^{12}$II.Physikalisches Institut, Justus Liebig Universit\"{a}t Giessen, 35392~Giessen, Germany}\\
\mbox{$^{13}$Department of Mechanical Engineering, Frederick University, 1036~Nicosia, Cyprus}\\
\mbox{$^{14}$Frederick Research Center, 1036~Nicosia, Cyprus}\\
\mbox{$^{15}$Faculty of Science, Palack\'{y} University Olomouc, 779 00~Olomouc, Czech Republic}\\
\mbox{$^{16}$Laboratoire de Physique des 2 infinis Irene Joliot-Curie, Universite Paris-Saclay, CNRS-IN2P3, F-91405~Orsay, France}\\
\mbox{$^{17}$Nuclear Physics Institute, The Czech Academy of Sciences, 25068~Rez, Czech Republic}\\
\mbox{$^{18}$Uniwersytet Warszawski, Instytut Fizyki Do\'{s}wiadczalnej, 02-093~Warszawa, Poland}\\
\mbox{$^{19}$Warsaw University of Technology, Faculty of Physics, 00-662~Warsaw, Poland}\\
\mbox{$^{20}$Bergische Universit\"{a}t Wuppertal, Department of Physics, 42119~Wuppertal, Germany}\\
\\
\mbox{$^{a}$ also at Helmholtz Research Academy Hesse for FAIR (HFHF), Campus Giessen, ~Giessen, Giessen}\\
\mbox{$^{b}$ also at Instituto Politecnico de Coimbra, Instituto Superior de Engenharia de Coimbra, 3030-199~Coimbra, Portugal}\\
\mbox{$^{c}$ also at Doctoral School of Exact and Natural Sciences, Jagiellonian University, ~Cracow, Poland}\\
\mbox{$^{d}$ also at Helmholtz Research Academy Hesse for FAIR (HFHF), Campus Darmstadt, 64390~Darmstadt, Germany}\\
\mbox{$^{e}$ also at Technische Universit\"{a}t Dresden, 01062~Dresden, Germany}\\
\mbox{$^{f}$ also at Helmholtz Research Academy Hesse for FAIR (HFHF), Campus Frankfurt, 60438~Frankfurt am Main, Germany}\\
\mbox{$^{g}$ also at Charles University, Faculty of Mathematics and Physics, 12116~Prague, Czech Republic}\\
\mbox{$^{h}$ also at Czech Technical University in Prague, 16000~Prague, Czech Republic}\\
\mbox{ e-mail: hades-info@gsi.de (J.~Stroth)}\\
}

\date{Received: date / Accepted: date}

\maketitle

\begin{abstract}
    Two-photon correlations can provide unique femtoscopic information, complementing those obtained with charged particles. They also allow to estimate the direct photon yield in the low $p_{T}$ region. 
    We present experimental results on $\gamma-\gamma$ correlations in Ag+Ag collisions at $\sqrt{s_{NN}}$ = 2.55 GeV, collected by the HADES experiment. Photons were measured using an electromagnetic calorimeter (ECAL) and two-photon correlations were determined as one-dimen-sional  functions $CF(q_{inv})$. We discuss the observed structures, along with their plausible origin. Finally, we estimate an upper limit for the correlation strength of direct photons for two confidence levels, resulting in values of the order $10^{-4}-10^{-3}$ .
\end{abstract}


\section{Introduction}
\label{sec:intro}

\subsection{From HBT to femtoscopy}

Hanbury Brown and Twiss intensity interferometry \cite{hbt} (often referred to as HBT), developed in the 1950s, was used to measure the angular sizes of stars via coincidental measurements of photons. Measuring the intensities of photons emitted from two distant points of the object of interest with two independent detectors separated by a certain distance, allows to determine the correlation between them. By varying the detectors separation, the angle between emission points can be estimated and, in consequence, the physical size of an object \cite{baym}. Thus, it can be shown that the correlation function measures the Fourier transform of the source distribution.

It was later discovered that a very similar principle can be used to probe properties of the particle emitting zone, as given by the position of their last interaction in nucleus-nucleus collisions. Femtoscopy \cite{Lisa,Marques}, originating from HBT measurements, can extract information about space-time characteristics of the particle-emitting sources (often referred to as fireball) created in the relativistic heavy-ion collision. The correlation function used in femtoscopy is defined by the Koonin-Pratt formula \cite{Koonin-pratt}:
\begin{equation}
    CF(q) = \int S(\mathbf{r})\left|\Psi(q,\mathbf{r}) \right|^2 d\mathbf{r}^3
    \label{eq:cf1}
\end{equation}
where $q = \left| \mathbf{p_{1}} - \mathbf{p_{2}}\right|$ is the four momentum difference between the two particles under consideration, $S(\mathbf{r})$ is the source function, describing the distribution of particle emission in space-time, and $\Psi(q,\mathbf{r})$ is the two-particle wave function. By knowing $\Psi(q,\mathbf{r})$, one can conclude on the source function, in particular on its length of homogeneity \cite{Lisa}, usually quantified by a radius parameter. The relative wave function is determined by the quantum statistics (in the case of identical particles, being Bose-Einstein or Fermi-Dirac  depending on their type) and the final state interactions between them (electromagnetic and/or strong).

The correlation function is usually normalized in such a way that $ CF(q) = 1$ implies no correlation. From the experimental point of view, as $S(\mathbf{r})$ is
not directly accessible, the statistical approach is used:
\begin{equation}
    CF(q) = \frac{P( \mathbf{p_{1}} , \mathbf{p_{2}})}{P( \mathbf{p_{1}})P( \mathbf{p_{2}})} = N\frac{\text{Same}(q)}{\text{Mixed}(q)}
    \label{eq:cf2}
\end{equation}
where $P( \mathbf{p})$ is the probability of observing particle with given momentum,  $P( \mathbf{p_{1}}, \mathbf{p_{2}})$  is the probability of observing a pair of particles, Same$(q)$ is the $q$ distribution constructed from particles taken from the same events, Mixed$(q)$ consists of those taken from different (mixed) events and $N$ is normalization constant. The mixed events must be as kinematically similar as possible to suppress non-femtoscopic contributions to the correlation function. The value of $N$ is adjusted so that $ CF(q) = 1$ is fulfilled in the region where no correlation is present. In the presented analysis, the $q$ range used for normalization is $(250;300)$ MeV/c.

\subsection{Importance of photons and classification}
\label{ssec:imp}
Photons are known to be valuable probes to study the space and time evolution of heavy-ion collisions, due to their lack of sensitivity to final state interactions, thus having relatively long mean free path \cite{DavidGabor,EMprobes}, which allows them to reach the detectors nearly undisturbed. As a result, their kinematic properties are well preserved until detection in the experimental setup. Additionally, photon emission is continuous through the collision's evolution, allowing one to probe the source characteristics before the kinetic freeze-out era. 

\begin{figure}
    \begin{minipage}{\columnwidth}
        \centering
        \includegraphics[width=\columnwidth]{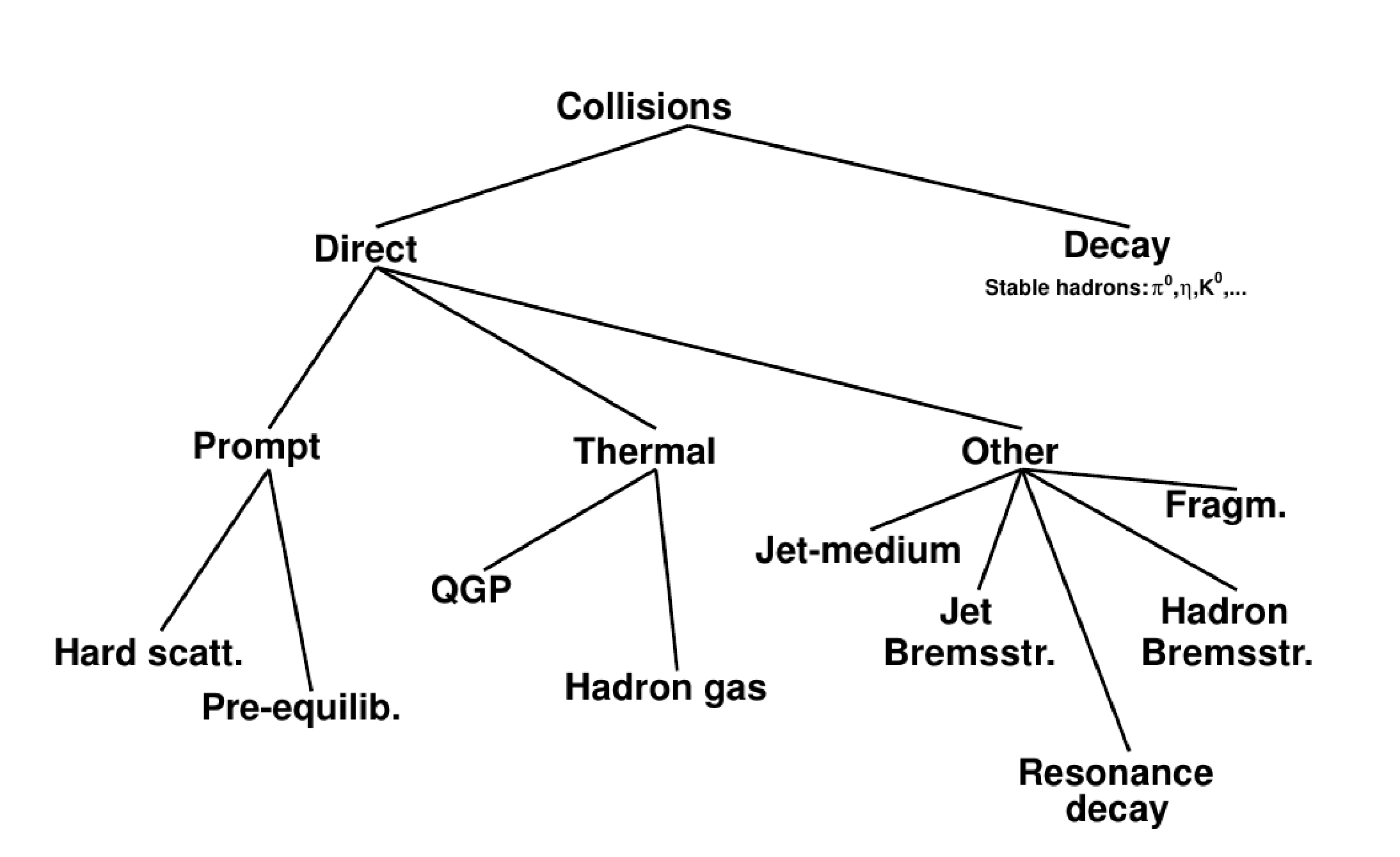}
    \end{minipage}
    \caption{Photon classification by origin, taken from Ref. \cite{DavidGabor}.}
    \label{fig:photonKinds}
\end{figure}

Photons can be divided into two groups according to their origin: electromagnetic decay and direct production from many different sources, see Fig. \ref{fig:photonKinds}. 

The decay photons originate from the decays of hadrons with a lifetime considerably larger than that of a fireball. In experimental analyses $\pi^{0}$ and $\eta$ mesons are regarded as the primary decay photon sources. 

Direct photons, on the other hand, can be further divided into prompt (originating from initial hard parton scattering), thermal (either from an equilibrated quark-gluon plasma or hadron gas), and other processes (jet-related interaction, \newline bremsstrahlung, etc.). Depending on the beam energy, the proportions of those sources will vary. In case of low-energy experiments (like HADES), some highly energetic sources (like QGP or jets) may not be present. Hence, despite a wide range of origins, direct photons have substantially lower yields than decay ones.

Experimentally, it is nearly impossible to distinguish between the different types of direct photons. In high multiplicity environments, the separation between direct and decay photons becomes extremely difficult due to a high combinatorial background (mostly uncorrelated photons from $\pi^0$ decays). Thus, an analysis of photons presents many challenges. 
In particular, it is very difficult to access direct photons at low transverse momenta via conventional methods. Here, the femtoscopy technique can be a valuable alternative approach.

\subsection{Photon femtoscopy}
The experimental correlation function for a pairs of identical uncharged bosons, following only quantum statistics and assuming a Gaussian source distribution, can be described by the parametrization \cite{Alexander1994}:
\begin{equation}
    CF(q) = 1 + \lambda e^{-R^2q^2}
    \label{eq:cf3}
\end{equation}
where $\lambda$ denotes the correlation strength and $R$ the femtoscopic radius. From Eq.~\eqref{eq:cf1} it follows that $CF(q\rightarrow0) = 2$, since $\left|\Psi(q,\mathbf{r}) \right|^2 = 1 + \cos(q\Delta\mathbf{r})$, hence the $\lambda$ parameter in Eq.~\eqref{eq:cf3} would be equal to unity. For photons, however, spin must be taken into account, since only photons with the same spin polarization contribute to Bose-Einstein correlations \cite{photonSpin}. Because of that, the effect of quantum statistics is reduced, resulting in $\left|\Psi(q,\mathbf{r}) \right|^2 = 1 + \frac{1}{2}\cos(q\Delta\mathbf{r})$ yielding $CF(q\rightarrow0) = 1.5$ and thus $\lambda = 0.5$.

Because the two-boson wavefunction depends both on $q$ and $\mathbf{r}$, one can employ the femtoscopy technique as a way to distinguish decay and direct photon signals. The emission points of direct and decay photons are separated by at least a hadron lifetime \cite{dimitri}, which implies that almost no measurable correlations can exist between them.
Residual correlations of decay photons can distort the shape of the correlation function \cite{dimitri}, however for $q < 50$~MeV/$c$ any enhancement should come only from direct photons. Nonetheless, direct photon correlation might be highly suppressed due to the much larger amount of decay photons. Therefore, the value of the $\lambda$ parameter will be strongly reduced. Knowing the total number of photons produced in a single event, $N_{\gamma, \; all}$, and the correlation strength of direct photons $\lambda_{direct}$, one can estimate the average direct photon yield $N_{\gamma, \; direct}$ with the formula \cite{wa98}:
\begin{equation}
    \lambda_{direct} = \frac{1}{2} \left( \frac{N_{\gamma, \; direct}}{N_{\gamma, \; all}} \right)^{2}
    \label{eq:yield}
\end{equation}

Previous studies on photon femtoscopy \cite{wa98,piasecki,STARdirectFemto,ALICEfemto,phoenix, TapsFemto} have shown that, regardless of collision energy and detector setup, such measurements pose a great challenge. Nevertheless, most of these studies have observed an HBT-like signal in the correlation functions. Due to limited statistics and/or large uncertainties, drawing a firm physical conclusion has always been challenging.

Here, we present a first measurement of the two-photon correlation function in Ag+Ag collisions at $\sqrt{s_{NN}} = 2.55$~GeV as performed by the HADES collaboration. The paper is organized as follows: first, we discuss the experimental setup, then the event selection, along with the selection of single particles and particle pairs. We then present the obtained correlation functions, along with the applied corrections and systematic analysis. We conclude with a discussion of the obtained results and their interpretation. 

\section{Experimental setup}
\label{sec:expSetup}
The High Acceptance Di-Electron Spectrometer (HADES) \cite{hades}, as part of the Facility for Antiproton and Ion Research (FAIR) phase-0 program \cite{fairPhase0}, studies the QCD phase diagram at low temperatures and high baryon chemical potentials. This region is characterized by conditions similar to those found in neutron star mergers \cite{HADESnature}.

HADES was optimized to operate at center-of-mass energies of a few GeV. Being a fixed target experiment, it covers almost the full azimuthal angle and polar angle within $18^\circ < \theta < 85^\circ$. Each subdetector is split in the azimuthal plane into six symmetrical sectors. The most important parts of the spectrometer are:

\begin{itemize}
    \item Target: contains 15 identical silver disc plates, having $\sim$2 mm radius and 40 $\mu$m thickness.
    \item START-VETO system: combination of CVD\footnote{Chemical Vapor Deposition} diamond detectors, containing two active planes. It registers and evaluates the amount of incoming beam particles before and after the target, thus the occurrence of reactions in the target and the start time of the collision (later referred to as $T0$), with a time resolution of 50~ps.
    \item Ring-Imaging Cherenkov Detector (RICH): provides electron and positron identification over a large momentum region. 
    \item Four layers of Multi-wire Drift Chambers (MDC): a set of detectors with a superconducting, toroidal magnet placed in between the first and last two MDC layers. The setup is used for track reconstruction and momentum determination, as well as for particle identification through measurement of the energy loss in the detector gas. 
    \item Resistive Plate Chamber (RPC)/ Time Of Flight (TOF): timing detectors placed after the last layer of MDC, covering $18^\circ < \theta < 45^\circ$ (RPC) and $44^\circ < \theta < 85^\circ$ (TOF).
    \item  Electromagnetic Calorimeter, see description in Sect~\ref{ssec:ecal}.
\end{itemize}

Events collected by the spectrometer for Ag+Ag collisions at $\sqrt{s_{NN}} =$ $2.55$~GeV, with a mean beam rate (number of beam particles per second) of $1.5 - 3.5$~MHz, were used in this analysis.

\subsection{Electromagnetic Calorimeter of HADES}
\label{ssec:ecal}

The electromagnetic Calorimeter (ECAL) \cite{ecal,ecal2} is located behind the RPC, covering polar angles of $12^\circ < \theta < 45^\circ$, and is designed for the detection of photons and di-electrons. It consists of 978 lead glass modules and is divided into six sectors. Each module contains a lead glass block of $420$~mm length (16.7 radiation lengths) and $92 \times 92$~mm$^2$ transverse area. It has an energy resolution of  $dE/E = 5.9 \,\% \,/ \sqrt{E \;\textrm{[GeV]}}$
and a Molière radius of $3.6$~cm. Particles hitting the calorimeter can induce an electromagnetic shower in the material. Emitted Cherenkov photons are then collected by photomultipliers (PMT). For electrons, positrons and photons, the response is proportional to the particle's energy. Hadronic showers are usually smaller, and the resulting signal has a defined value.

For high-energy particles the electromagnetic shower can extend over more than one module. In this case, neighboring activated modules are combined into clusters, for each of which the total deposited energy and the average time of arrival, as well as the position, of the incident particle is calculated. Each cluster undergoes a matching procedure to charged particles, quantified by a $N \sigma$ value defined as:
    \begin{equation}
        N\sigma = \sqrt{ \left( \frac{\Delta x}{\sigma_x} \right)^2 + \left( \frac{\Delta y}{\sigma_y} \right)^2} \;,
        \label{eq:mq}
    \end{equation}
where $\Delta x$, $\Delta y$ are the differences between the reconstructed position of the cluster in the calorimeter and the position of the charged particle track, projected onto the ECAL plane, and $\sigma_x$,$\sigma_y$ are the spatial resolutions of the ECAL. 
During the Ag+Ag beam in 2019 only sectors 2 and 4, as well as parts of sector 5 and 1 were operational.

\section{Analysis}
\label{sec:analysis}

In addition to the data collected by the spectrometer, simulations were conducted using the SMASH model \cite{Smash, SmashSoftware} and the PLUTO event generator \cite{pluto}, combined with a detailed simulation of the detector response modeled in the HGeant package (based on GEANT 3.21 \cite{geant}). With dedicated digitization algorithms it reproduces the resolution and acceptance of the real detector, using the same data structures as for real collisions. Due to that both data sources (simulated and real data) use the same reconstruction procedure.
 
It should be emphasized that simulations do not include quantum statistical or final-state interaction effects. Their sole purpose is to serve as a benchmark for detector effects, allowing for the development of necessary corrections.

\subsection{Event selection}
\label{ssec:eventSelection}

The events selected for the analysis must fulfill certain criteria, including:
\begin{itemize}
    \item having an event vertex (point of minimum distance of all tracks from the beam axis) reconstructed within one of the 15 target plates.
    \item no additional hits in the START-VETO system within a window of $15$~ns beginning with the time of the collision (T0), assuring its correct estimation.
    \item having a correlated VETO hit within $\pm2$ ns window to any other START hit, occurring within $15-350$ ns from the one selected for establishing T0.
    \item exhibiting a number of charged particle hits in the RPC and TOF detectors between 38 and 173, corresponding to the $0-40$~\% most central events. 
\end{itemize}
The detailed description of the event selection criteria can be found in \cite{Simon}.

The centrality is estimated by measuring charged particle hits in the TOF and RPC detectors, which are later compared with predictions from the Glauber Monte-Carlo model \cite{HADEScentrality}. The selected centrality range translates into the impact parameter range of $0 < b < 7.7$~fm. Additionally, only events with at least one photon candidate reconstructed were considered, resulting in total in $1.23 \:\cdot\: 10^9$ analyzed events, which is roughly $35$~\% of all recorded events.

\subsection{Photon reconstruction}
\label{ssec:photonReco}
Within the HADES experiment, photons can in principle be reconstructed using either the information provided by the ECAL or by the photon conversion method (PCM \cite{pcm}). 

\begin{figure}
    \begin{minipage}{\columnwidth}
        \centering
        \includegraphics[width=\columnwidth]{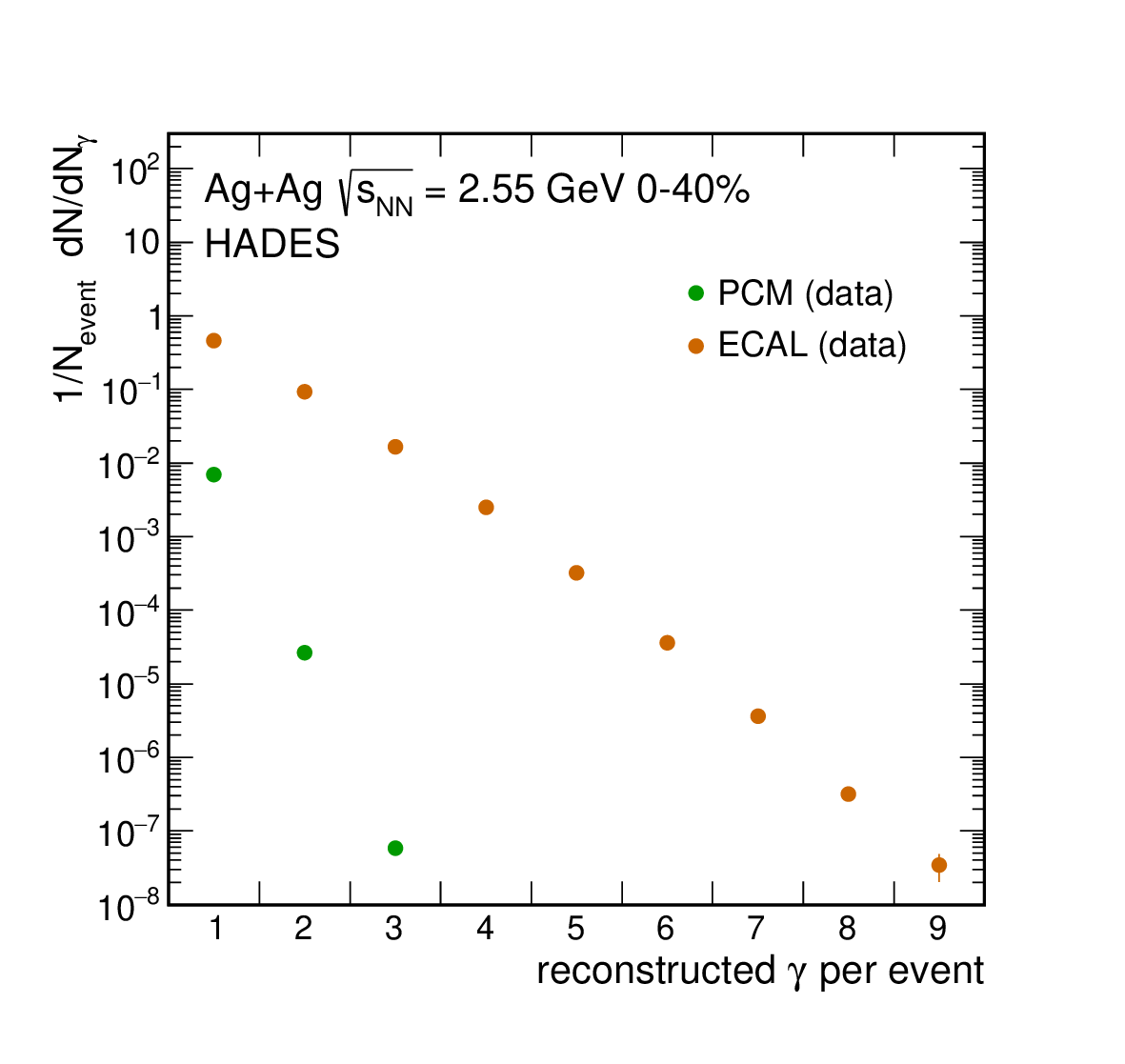}
    \end{minipage}
    \caption{Comparison of the measured number of photons per event between the photon conversion method (PCM) and ECAL in Ag+Ag collisions at $\sqrt{s_{NN}} = 2.55$~GeV.}
    \label{fig:photonPerEv}
\end{figure}

Calorimetry has the advantage of providing a high reconstruction efficiency; however, it has some limitations due to the finite granularity of the detector modules, caused by the transverse size of the modules, resulting in a moderate angular resolution. Additional limiting factors may include imperfections in the clusterization procedure as well as the energy resolution of the calorimeter, which deteriorates towards low photon energies. 
Alternatively, PCM benefits from the fact that HADES was specifically built for the detection of dielectrons and therefore has a high $e^{\pm}$ reconstruction efficiency. This, in theory, should allow for the reconstruction of photons via the $\gamma\rightarrow e^{+}e^{-}$ channel. However, as the HADES's primary physics program are leptons from light vector meson and virtual photon decays, the detector was designed with a very low material budget \cite{pcm}. Thus photon conversions in the detector material are suppressed in order to reduce the background in the dielectron measurements. As a consequence, the efficiency of the $\gamma\rightarrow e^{+}e^{-}$ channel was found to be insufficient for the studies intended here (the probability of reconstructing two $\gamma$ per event is only $\sim 10^{-5}$, see Fig. \ref{fig:photonPerEv}). In addition, the signal purity suffers from a combinatorial background in the measured $e^{\pm}$ pairs.
Therefore, we restrict this analysis to photons measured with ECAL only.

\begin{table}
\centering
\caption{Photon selection criteria}
\label{tab:photonSelection}
\begin{tabular*}{\columnwidth}{@{\extracolsep{\fill}}ll@{}}
\hline
Variable & Criterion\\
\hline
Total cluster energy & $> 100$~MeV \\
Minimum cell energy & $> 50$~MeV \\
$N \sigma$ matching quality to charged track & $>6$ \\
Opening angle to any charged track & $> 2.8^{\circ}$ \\
Velocity ($\beta$) & 1 $\pm \sigma$ \\
Cluster size & $<4$ \\
\hline
\end{tabular*}
\end{table}

The photon candidates have to fulfill the criteria presented in Tab.~\ref{tab:photonSelection}. The $100$~MeV energy threshold was used to suppress the contamination by fast neutrons, as they cannot be distinguished from photons by any other detector and most of neutrons have their energy deposition in the ECAL below this value. The $50$~MeV restriction on the minimum cell energy is motivated by the dependence of the calorimeter resolution proportional to $\sim 1\,/\,\sqrt{E}$ and thus limits the resolution to values better than $\sim 25$~\%. The $N \sigma$ criterion ensures that ECAL clusters are isolated, i.e. that no matches to charged particle tracks are found. To account for cases where RPC clusters might not be correctly assigned to charged tracks, a minimum opening angle between each ECAL cluster and a reconstructed particle candidate is imposed. Following simulation results, $2.8^\circ$ was found to be a suitable value. To suppress the contribution of particles which were not rejected by the $N \sigma$ or opening angle criteria (mostly remaining fast neutrons and $\pi^\pm$), the particle velocity must be around $\beta = 1$ within $1 \sigma$ (estimated separately for each module of ECAL). Finally, only clusters containing fewer than four modules are considered as valid, since larger clusters are almost exclusively composed of more than two different ones, which were falsely combined into a single structure. 

\begin{figure*}
    \begin{minipage}{0.5\linewidth}
        \centering
        \includegraphics[width=\columnwidth]{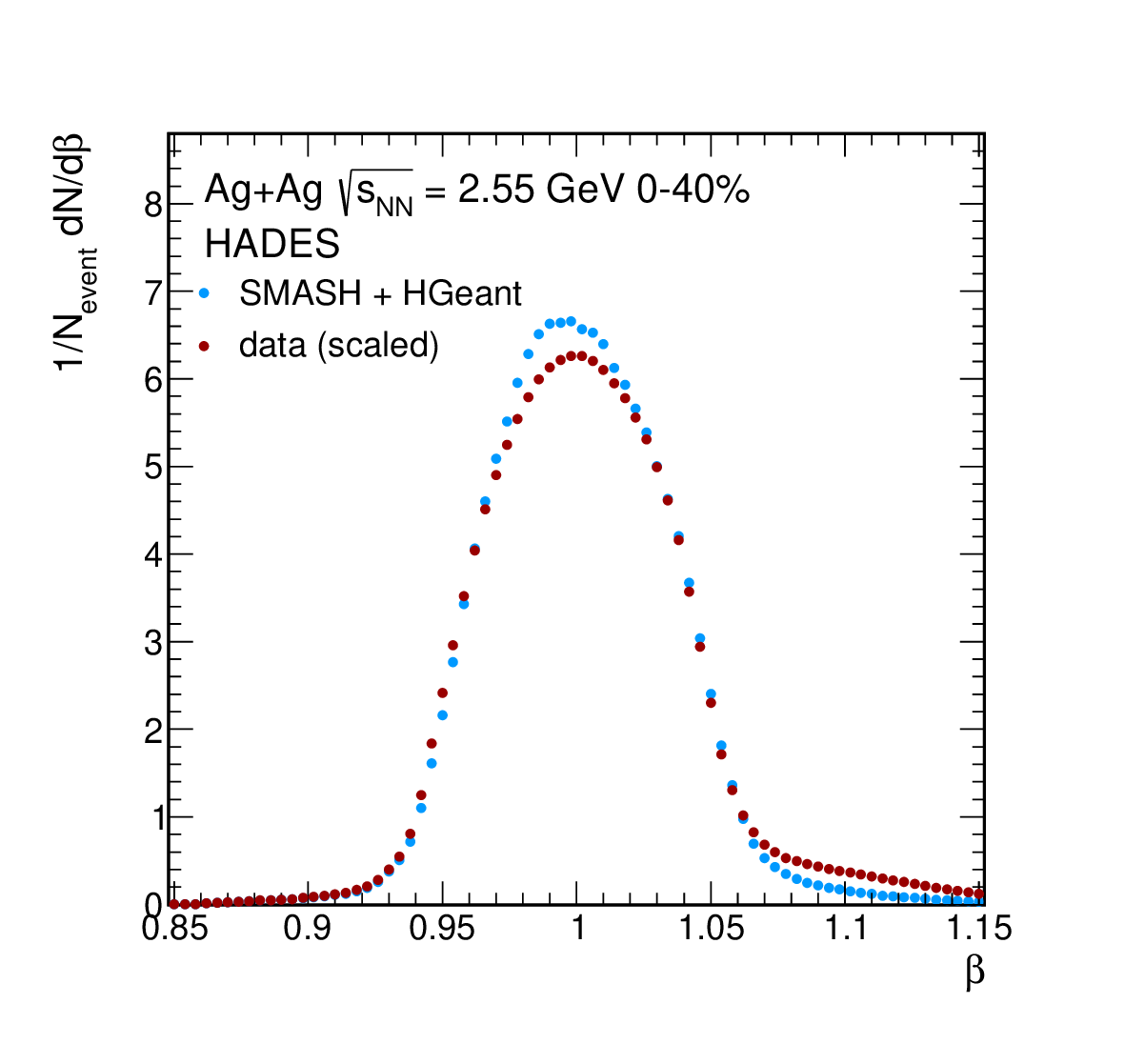}
    \end{minipage}
    \begin{minipage}{0.5\linewidth}
        \centering
        \includegraphics[width=\columnwidth]{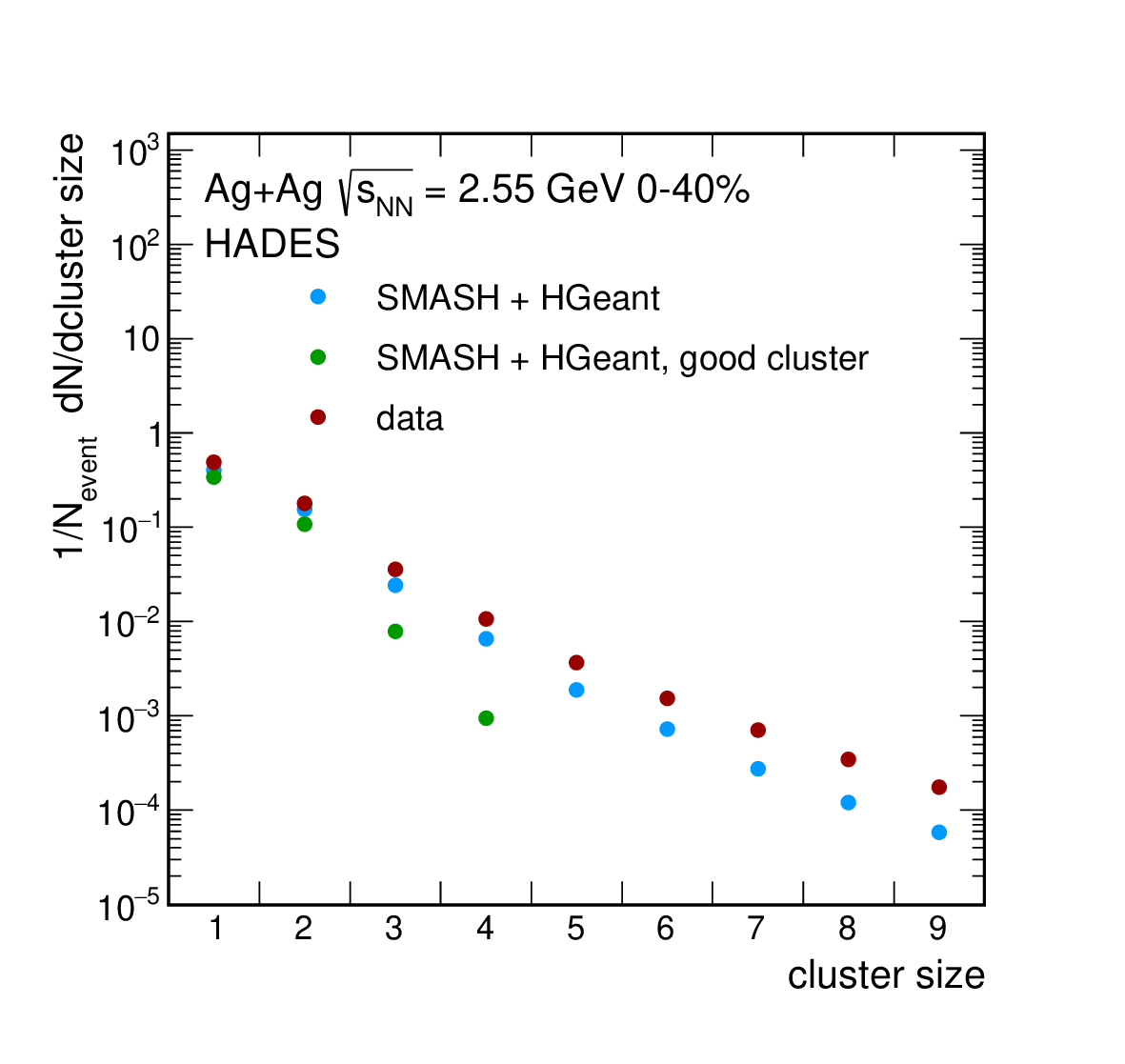}
    \end{minipage}
    \caption{Left panel - $\beta$ distribution of accepted photon candidates in real data (red symbols) compared with simulation results (blue symbols). The experimental data points were scaled down to match the integral of simulation distribution for easier shape comparison.
    Right panel - cluster size (number of triggered modules) distributions of accepted photon candidates. Shown are measurements (red symbols), as well as simulation results for all clusters (blue symbols) and for clusters matched to photons, with all modules properly assigned ("good cluster", green symbols).} 
    \label{fig:betaSize}
\end{figure*}

A comparison of the cluster size distributions for the reconstructed photon candidates is presented in the right panel of Fig.~\ref{fig:betaSize} for real and simulated data. A fair agreement is observed up to a cluster size of nine (red and blue symbols), with the real data consistently above the simulated ones, due to additional photon sources in real data in comparison to simulations. In addition, the simulation allows for selecting clusters composed of modules triggered by the same photon (here labelled "good cluster", green symbols), thereby rejecting merged clusters and those triggered by other particles. This distribution drops much faster than the inclusive distribution, indicating that clusters with size above three are heavily dominated by non-photonic origins or false reconstruction.

The left panel of Fig.~\ref{fig:betaSize} shows the velocity distribution of the accepted photons, both for simulation (blue symbols) and real data (red symbols). The simulation was scaled up to match the integral of real data. 

\subsection{Close track effects}
\label{ssec:TwoTrack}
 The correlation functions of photon pairs are studied as a function of the one-dimensional momentum difference $q_{inv}$. For massless particles it is equal to the invariant mass of the pair $M_{\gamma\gamma}$:
\begin{eqnarray}
        q_{inv} & = & \sqrt{\left| \mathbf{p_1} - \mathbf{p_2}\right|^2 - \left( E_1 - E_2 \right)^2} \nonumber \\
                & = & \sqrt{2 \, E_1 E_2 \, (1 - \cos \alpha_{\gamma\gamma})} \nonumber \\
                & = & M_{\gamma\gamma}
        \label{eq:qinv}
\end{eqnarray}
where $\mathbf{p}$ is the photon momentum, $E$ its energy and $\alpha_{\gamma\gamma}$ the opening angle between two photons.

Before creating a correlation function, close track effects need to be addressed, as well as the influence of pair purity and the detector's resolution. Reconstruction errors can lead to both enhancement and suppression of pairs at low $q_{inv}$ \cite{STARMergeSplit}, which is the region of interest for any femtoscopic measurement. The enhancement caused by falsely reconstructing a single particle as a pair with a very small relative angle is often referred to as splitting. The opposite effect, called merging, can occur if a pair of particles with a very low relative angle is reconstructed as a single particle, causing an abnormal decrease of the correlation function for small relative momenta.

The detection of two photons close in momentum and configuration space is limited by the transverse area of the ECAL modules and the imposed lower limit of their energy (100 MeV). Consequently, a photon pair with $\alpha_{\gamma\gamma} < 4.2^{\circ}$ cannot be separated within the same event. Additionally, in rare cases (0.1\% of all recorded photons), the clusterization algorithm may falsely reconstruct one photon as two clusters with an opening angle near the $4.2^{\circ}$ threshold. This might be caused by an electromagnetic shower starting in the non-active material of the RPC detector, just before the ECAL modules. Despite its rarity, split clusters were found to populate almost exclusively the region of very low $q_{inv}$,  significantly affecting the shape of the correlation function.  

Due to the detector dead time, each ECAL module can be triggered only once per event, thus detect only one photon. This in turn suppresses photon pairs with small spatial separation within the same event. Such a limit does not apply when we consider the mixed event technique, as photons from different events can trigger the same module, allowing for pairs with inaccessible opening angle (below $4.2^{\circ}$ threshold). Because of that there is a difference in the pair angular acceptance between same and mixed events, leading to the merging effect in correlation functions, if not taken into account properly.

As the two mentioned effects (splitting and merging) overlap, they can be treated simultaneously with appropriate limitations on $\alpha_{\gamma\gamma}$. From Eq.~\eqref{eq:qinv} it follows that high caution is required when setting such criteria, since it directly limits the range of $q_{inv}$ for the particle pairs. Using simulations, the minimally allowed opening angles were determined. These values were estimated using opening angle distributions calculated for the same and mixed events, taking into consideration single and multiple clusters reconstructed from the same photon. It was observed that in the case of multiple reconstruction the distribution was highly biased towards very low values. 

\begin{figure}
    \begin{minipage}{\columnwidth}
        \centering
        \includegraphics[width=\columnwidth]{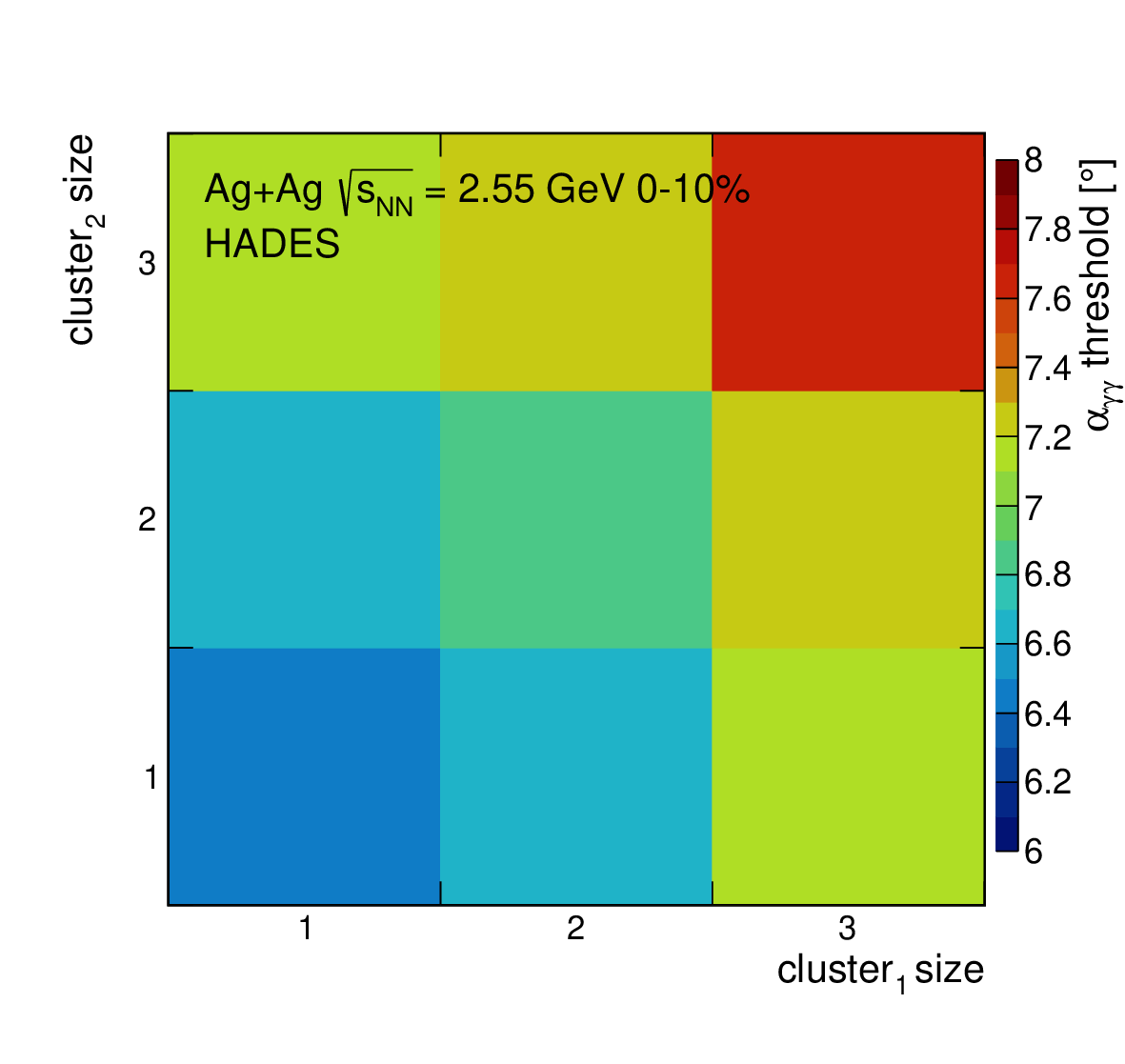}
    \end{minipage}
    \caption{Matrix of minimum allowed opening angle ("$\alpha_{\gamma\gamma}$ map") as a function of the cluster sizes of both photon candidates, for the $0 - 10$~\% centrality class. See text for definitions and details.}
    \label{fig:map}
\end{figure}

 For each cluster size combination, the smallest $\alpha_{\gamma\gamma}$ unifying the detector acceptance between the same and mixed events, and rejecting as much as possible the contribution of split clusters was determined. The values of lower threshold on $\alpha_{\gamma\gamma}$ were collected in matrices of opening angle thresholds (later referred to as "$\alpha_{\gamma\gamma}$ maps"). Separate maps were prepared for each $10$~\% interval of collision centrality, since differences were observed for different centrality classes. This approach limits the impact of splitting and merging effects, preserving a sufficiently large sample of photon pairs. For $q_{inv} < 50$~MeV/$c$ it reduces the split pair contributions by 40\% for both clusters with size = 1, up to 70\% for clusters with size = 3, simultaneously reducing the total pair quantity by less than 4\%. Fig.~\ref{fig:map} shows the $\alpha_{\gamma\gamma}$ map for the $0 - 10$~\% centrality class.

Besides splitting and merging, the detector resolution and the photon sample purity can impact the correlation function. The first one causes a small distortion of the reconstructed momenta compared to those carried by the original particles, resulting in distortion of the correlation function at low $q_{inv}$. The second one decreases the strength of the correlation, due to contamination by other particle species in the sample, which do not contribute to the correlation. To account for these effects, resolution and purity corrections were introduced, as described by Eqs.~\eqref{eq:resolution} and \eqref{eq:purity}. Both corrections were estimated using simulations:

\begin{equation}
    \begin{split}
    res(q_{inv}) & = \frac{CF_{MC-reco.}(q_{inv})}{CF_{MC-model}(q_{inv})} - 1\\
     CF_{res}(q_{inv})& = \frac{CF_{raw}(q_{inv})}{1 + res(q_{inv})} \;
    \end{split}
    \label{eq:resolution}
\end{equation}
where $res(q_{inv})$ is the resolution factor. $CF_{MC-model}(q_{inv})$ is the correlation function built from the Monte Carlo model simulation data. $CF_{MC-reco.}(q_{inv})$ is the correlation function built from simulated data after applying the full reconstruction chain and thus taking resolution effects into account. $CF_{raw}(q_{inv})$ is the measured correlation function. Finally, the $CF_{res}(q_{inv})$ represents the correlation function after correction for the detector resolution.

\begin{equation}
    \begin{split}
    pur(q_{inv}) & = \frac{N_{\gamma\gamma}(q_{inv})}{N_{all}(q_{inv})} \\
     CF(q_{inv})& = \frac{CF_{res}(q_{inv})-1}{pur(q_{inv})} +1 \; ,
    \end{split}
    \label{eq:purity}
\end{equation}
where $pur(q_{inv})$ is the purity of a $q_{inv}$ distribution, $N_{\gamma\gamma}(q_{inv})$ is the distribution of true photon pairs, $N_{all}(q_{inv})$ is the distribution of all accepted pairs and $CF(q_{inv})$ represents the fully corrected correlation function.

\begin{figure*}
    \begin{minipage}{0.5\linewidth}
        \centering
        \includegraphics[width=\columnwidth]{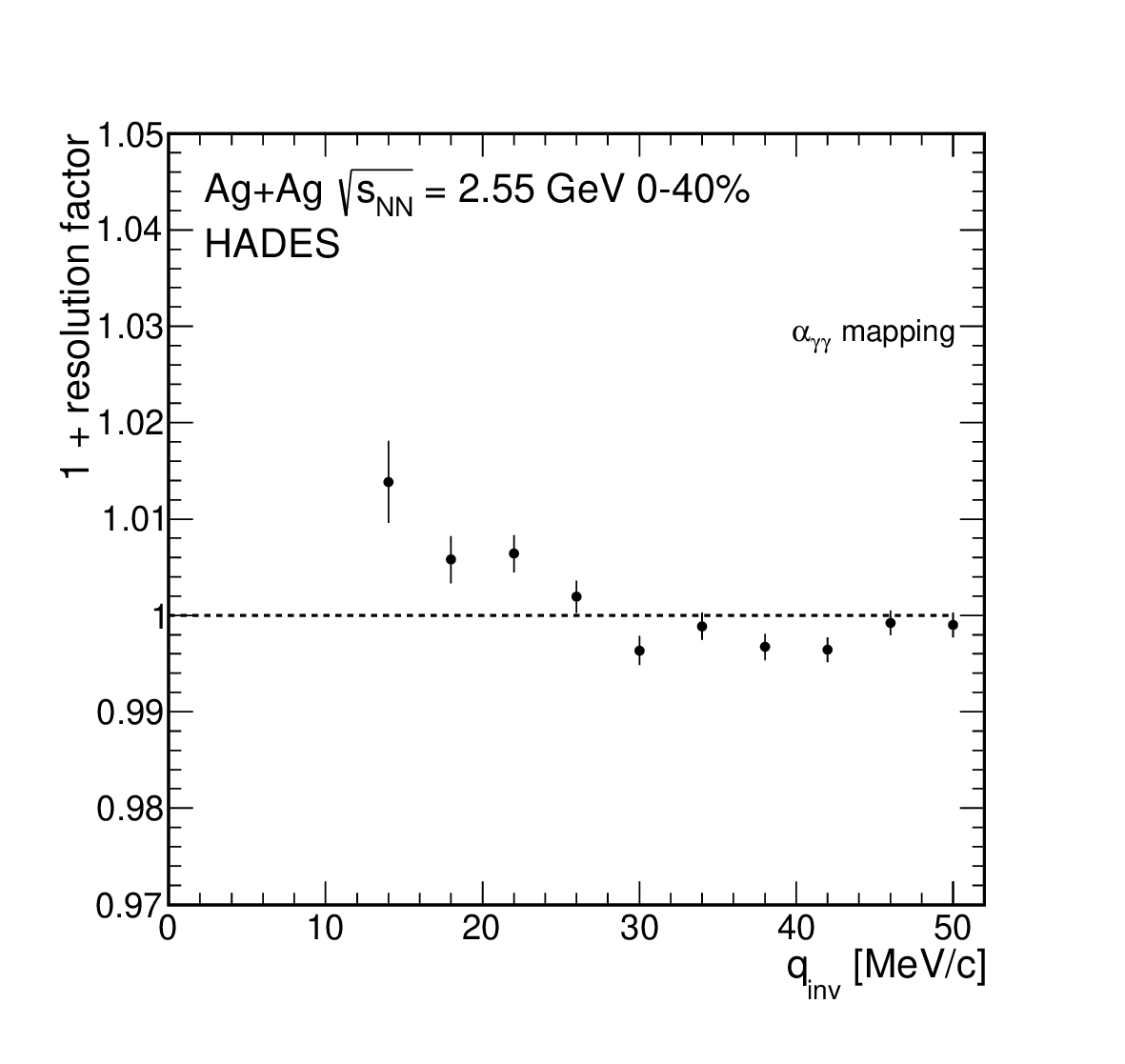}
    \end{minipage}
    \begin{minipage}{0.5\linewidth}
        \centering
        \includegraphics[width=\columnwidth]{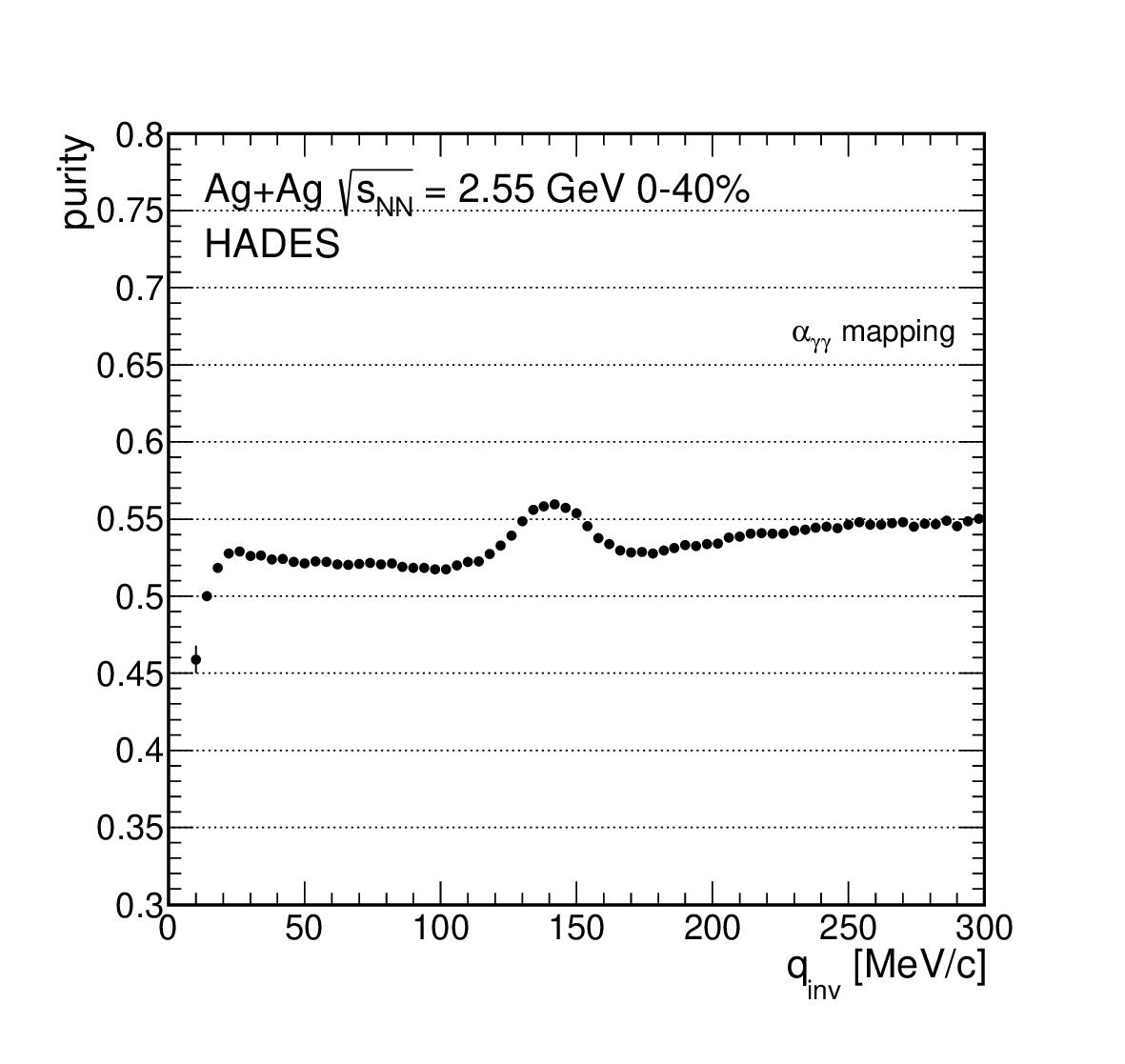}
    \end{minipage}
    \caption{Left panel - Momentum resolution factor $res(q_{inv})$, after applying the $\alpha_{\gamma\gamma}$ maps, as calculated using the SMASH model and HGeant. Right panel -  Purity of accepted pairs for the correlation function $pur(q_{inv})$, after applying the $\alpha_{\gamma\gamma}$ maps. See text for definitions and details.}
    \label{fig:purityAndResolution}
\end{figure*}

Figure~\ref{fig:purityAndResolution} shows the resolution (left panel) and the purity (right panel) as a function of $q_{inv}$.

\subsection{Systematic uncertainty analysis}
\label{ssec:systematics}
Systematic uncertainties were calculated by varying the values of selection criteria for both single particles and particle pairs. These variations are defined as:
\begin{align}
    \delta(q_{inv}) = CF_{def.}(q_{inv}) - CF_{var.}(q_{inv}) \\ 
    \Delta_{syst}(q_{inv}) = \sqrt{\sum_{i=1}^{N}\frac{1}{2}[\delta_{i+}(q_{inv})^{2} + \delta_{i-}(q_{inv})^{2}]}
\end{align}
where $CF_{def.}(q_{inv})$ is the correlation function obtained with the default criteria, $CF_{var.}(q_{inv})$ the one after applying varied criteria, $\delta_{i\pm}(q_{inv})$ represents the variation difference of a certain criterion, created by increasing/decreasing strength of the selection and $\Delta_{syst}(q_{inv})$ represents the total systematic uncertainty.

\begin{table}
\centering
\caption{Selection criteria variation}
\label{tab:variations}
\begin{tabular*}{\columnwidth}{@{\extracolsep{\fill}}lll@{}}
\hline
variable & default criterion & variation \\
\hline
$N \sigma$ matching quality & $> 6$ & $\pm$ 1 \\
Charged particle o.a & $> 2.8^{\circ}$ & $\pm$ $0.8^{\circ}$\\
Minimum cell energy & $> 50$ MeV & $\pm$ 10 MeV\\
$N \sigma$ $\beta$ & $< 1$ & $\pm$ 0.25\\
$\alpha_{\gamma\gamma}$ & value from matrix & $\pm$ 5\%\\
Resolution correction & value from a function & $\pm$ 10\%\\
Purity correction & value from a function & $\pm$ 20\%\\
\hline
\end{tabular*}
\end{table}

The default values for selection criteria, as well as variation values, are presented in Tab.~\ref{tab:variations}. Additionally, the Barlow test \cite{barlow} principle was utilized. It aims to reduce the influence of limited statistics on systematic uncertainties. Thus, each $\delta(q_{inv})$ component is modified by the statistical uncertainty:

\begin{align}
    CF_{diff}(q_{inv}) = CF_{def.}(q_{inv}) - CF_{var.}(q_{inv}) \\ \sigma_{diff}(q_{inv}) = \sqrt{|stat_{def.}(q_{inv})^{2} - stat_{var.}(q_{inv})^{2}|} \\
    \xi_{diff}(q_{inv}) = \sqrt{CF_{diff}(q_{inv})^{2} - \sigma_{diff}(q_{inv})^{2}}
\end{align}

\begin{equation}
    \delta(q_{inv}) = 
    \begin{cases}
         \xi_{diff}(q_{inv}) \:\:\:\text{if} \:\:\: CF_{diff}(q_{inv}) > \sigma_{diff}(q_{inv})\\
        0  \qquad\qquad\:\: \text{otherwise}\:,
    \end{cases} 
    \label{eq:syst}
\end{equation}
where $stat(q_{inv})$ is the statistical uncertainty of the correlation function (with default or changed criteria) and $\sigma_{diff}(q_{inv})$ represents the impact of the statistical uncertainties on the variation. Following Eq.~\eqref{eq:syst}, the variation factor $\delta(q_{inv})$ is set to zero, if the statistical uncertainty is larger than the difference caused by the variation, as in this case the impact of such a variation cannot be significantly determined. 

\subsection{Fitting the correlation functions}
\label{ssec:fit}

\begin{figure*}[t]
    \begin{minipage}{0.5\linewidth}
        \centering
        \includegraphics[width=\columnwidth]{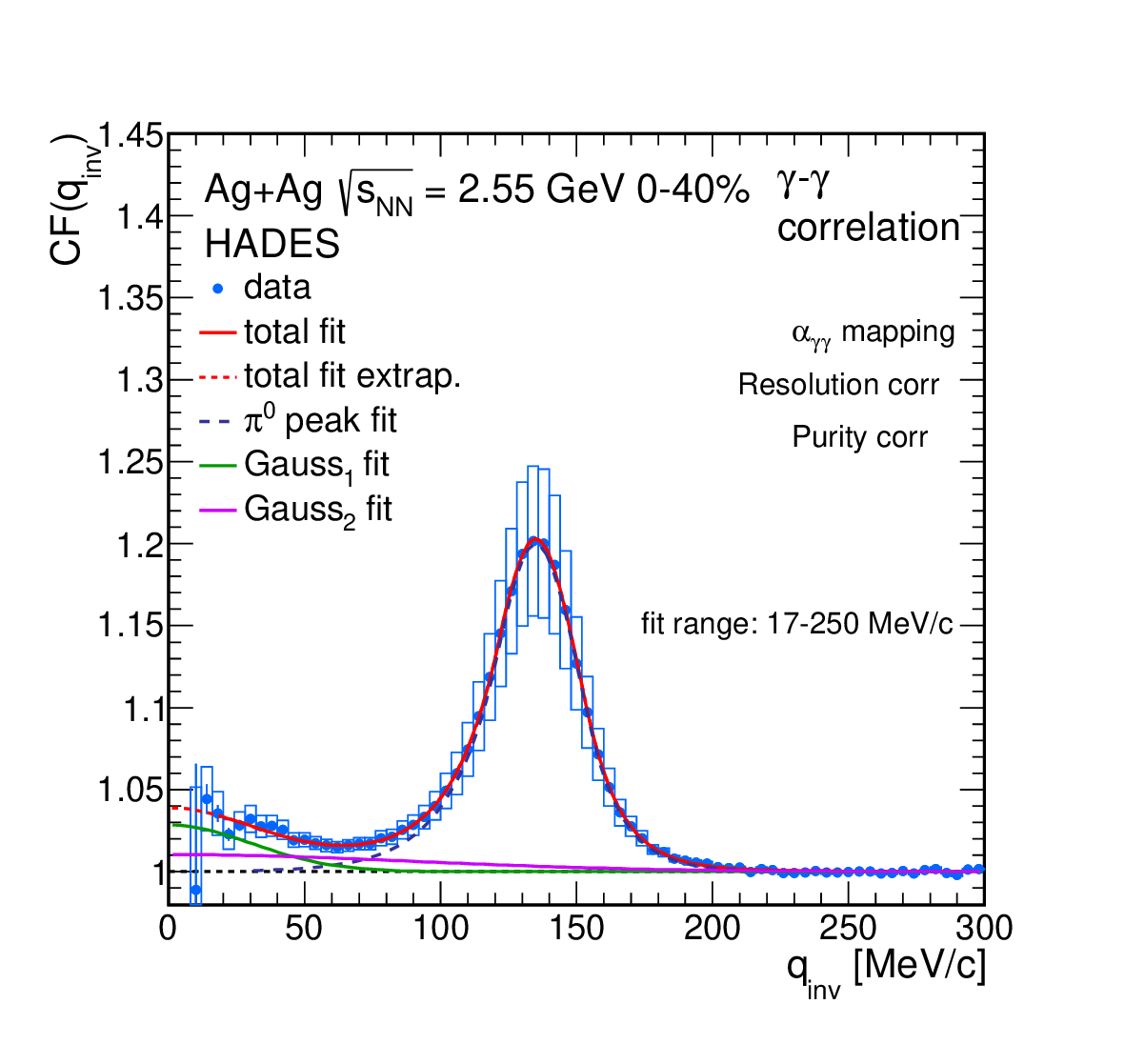}
    \end{minipage}
    \begin{minipage}{0.5\linewidth}
        \centering
        \includegraphics[width=\columnwidth]{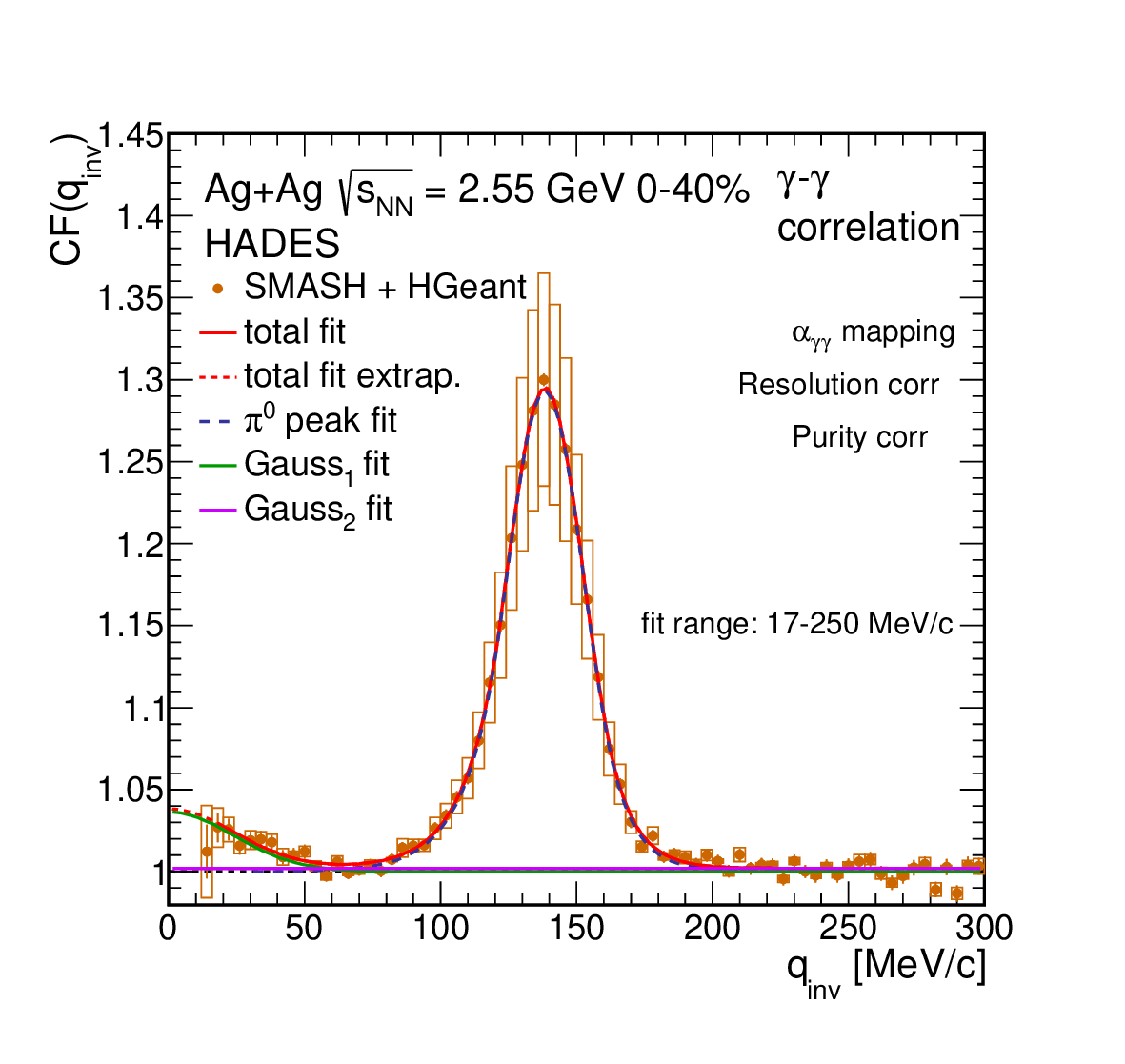}
    \end{minipage}
    \caption{$\gamma-\gamma$ correlation functions for $0 - 40$~\% most central Ag+Ag collisions at $\sqrt{s_{NN}} = 2.55$~GeV, after applying the $\alpha_{\gamma\gamma}$ maps, as well as the resolution correction and purity corrections, fitted with the formula given in Eq.~\eqref{eq:fitFunction}. Boxes around data points represent systematic errors. The solid red line represents the fitted region, and the dashed red line shows the extrapolated fit. The green and purple lines represent the single Gaussian components. The dashed dark blue line describes the $\pi^0$ peak. Left panel - real data. Right panel - SMASH+HGeant simulations (without quantum statistics effects).}
    \label{fig:CFfinal}
\end{figure*}

The final correlation functions, obtained from simulations and real data, were fitted with the formula:
\begin{equation}
\begin{split}
        CF(q_{inv}) = 1 + \lambda_{1}\,e^{-R_{1}^{2}\,q_{inv}^{2}} + \lambda_{2}\,e^{-R_{2}^{2}\,q_{inv}^{2}} \\ 
    +f_{\pi^0}(q_{inv},\mu_{\pi^0},\sigma_{\pi^0},k_{L},k_{H})
    \label{eq:fitFunction}
\end{split}
\end{equation}
The first term is related to direct photons correlation, while second is attributed to residual correlation of $\pi^0$-$\pi^0$. The $f_{\pi^0}$ term, describing the $\pi^0$ mass peak is defined as \cite{crystalball}:
\begin{equation}
    f_{\pi^0} = 
    \begin{cases}
        e^{\frac{k_{L}^{2}}{2}} + k_{L} \cdot \left( \frac{q_{inv} - \mu_{\pi^0}} {\sigma_{\pi^0}} \right) \quad\:\, \text{if} \quad \left( \frac{q_{inv} - \mu_{\pi^0}} {\sigma_{\pi^0}} \right) \leq -k_{L}  
        \\
         e^{\frac{k_{H}^{2}}{2}} - k_{H} \cdot \left( \frac{q_{inv} - \mu_{\pi^0}} {\sigma_{\pi^0}} \right) \quad \text{if} \quad \left( \frac{q_{inv} - \mu_{\pi^0}} {\sigma_{\pi^0}} \right) > k_{H}         
                \\
         e^{-\frac{1}{2}\left( \frac{q_{inv} - \mu_{\pi^0}} {\sigma_{\pi^0}} \right)^{2}}  \quad\quad\quad\:\: \text{otherwise} \:             
    \end{cases}
\end{equation}
where $\mu_{\pi^0}$ is the center of the $\pi^0$ peak and $\sigma_{\pi^0}$ its width. The shape parameters $k_{L}$ and $k_{H}$ define the low and high $q_{inv}$ regions of the peak tails, respectively.

The used function is a modified version of single Gaussian parametrization, as described by Eq.~\eqref{eq:cf3}. It allows to take into account the residual correlations, which cannot be excluded from the data, hence require additional parametrization factors. 

Fig.~\ref{fig:CFfinal} shows the final $\gamma-\gamma$ correlation functions from real data (left panel) and simulations (right panel), fitted with Eq.~\eqref{eq:fitFunction}. Boxes around the data points represent the systematic uncertainties. Parameters of both fits are presented in the Tab.~\ref{tab:parsExp} and ~\ref{tab:parsSim}.

\begin{table}
	\centering
	\caption{Obtained fit parameters for real data (Fig.~\ref{fig:CFfinal}, left panel)}
	\label{tab:parsExp}    
	\begin{tabular*}{\columnwidth}{@{\extracolsep{\fill}}lllll@{}}
		\hline
		parameter & value & stat. uncert. & syst. uncert. (Barlow) \\
		\hline
		$\chi^2/NDF$ & $2.11$\\
		$R_{1}$ [fm] & $4.59$ & $\pm 0.42$ & $+0.45\;-0.39$ \\
		$\lambda_{1}\cdot10^{2}$ & $2.85$  & $\pm 0.28$ & $+0.96\;-1.23$ \\
		$R_{2}$ [fm] & $1.55$ & $\pm 0.13$ & $+0.03\;-0.00$ \\
		$\lambda_{2}\cdot10^{2}$ & $1.04$ & $\pm 0.22$ & $+0.20\;-0.21$ \\
		$\mu_{\pi^{0}} $[MeV/$c^{2}$] & $135.04$ & $\pm 0.04$ & $+0.16\;-0.22$ \\
		$\sigma_{\pi^{0}} $[MeV/$c^{2}$] & $15.55$ & $\pm 0.06$ & $+0.14\;-0.19$ \\
		$k_{L}$ & $0.91$ & $\pm 0.02$ \\
		$k_{H}$ & $1.28$ & $\pm 0.02$ \\
		\hline
	\end{tabular*}
\end{table}

\begin{table}
	\centering
	 \caption{Obtained fit parameters for simulation (Fig.~\ref{fig:CFfinal}, right panel)}
	 \label{tab:parsSim}    
	 \begin{tabular*}{\columnwidth}{@{\extracolsep{\fill}}lllll@{}}
		 \hline
		 parameter & value & stat. uncert. & syst. uncert. (Barlow) \\
		 \hline
		 $\chi^2/NDF$ & $1.09$\\
		 $R_{1}$ [fm] & $5.82$ & $\pm 0.59$ & $+0.16\;-0.67$ \\
		 $\lambda_{1}\cdot10^{2}$ & $3.64$  & $\pm 0.70$ & $+1.04\;-1.12$ \\
		 $R_{2}$ [fm] & $0.00$ & $\pm 0.83$ & $+0.00\;-0.00$ \\
		 $\lambda_{2}\cdot10^{2}$ & $0.20$ & $\pm 0.09$ & $+0.08\;-0.05$ \\
		 $\mu_{\pi^{0}} $[MeV/$c^{2}$] & $138.42$ & $\pm 0.11$ & $+0.06\;-0.18$ \\
		 $\sigma_{\pi^{0}} $[MeV/$c^{2}$] & $14.22$ & $\pm 0.15$ & $+0.15\;-0.09$ \\
		 $k_{L}$ & $1.14$ & $\pm 0.04$ \\
		 $k_{H}$ & $1.53$ & $\pm 0.10$ \\
		 \hline
	\end{tabular*}
\end{table}

Most pronounced of the residual contributions (dashed dark blue line) is the peak centered at $\pi^0$ mass  ( $M_{\gamma\gamma} = q_{inv} \approx 135$~MeV/$c$), attributed to photons coming from  $\pi^0\rightarrow \gamma \gamma$ decay channel.
One can see the difference in its height and the slight offset of the position between real data and simulation. The height difference is most likely a consequence of additional photon sources in real data in comparison to simulations, which increase the combinatorial background in the same-event $q_{inv}$ distribution, thus lowering the magnitude of the peak. Additionally, the simulations might underestimate the $\pi^{0}$ yield. The offset is caused by the $\pi^0$ implementation in the SMASH model, having the same mass as $\pi^\pm$, while in reality they are $\approx 4.5$~MeV/$c^2$ apart. The fitted function describes this structure well, yielding values for $\mu_{\pi^0}$ in line with expectations and $\sigma_{\pi^{0}}$ consistent with the nominal value \cite{ecal3}. 

The second residual contribution (solid purple line) can only be observed in the case of real data, where the correlation function does not reach unity before the $\pi^{0}$ peak, but rather at around $q_{inv} \approx 200$~MeV/$c$. This extra enhancement has been observed in previous studies \cite{phoenix} and described by theory \cite{dimitri}, as being most likely a remnant of $\pi^0-\pi^0$ femtoscopic correlations. Such a component is not present in case of simulations, because, as mentioned in Sect.~\ref{sec:expSetup}, no femtoscopic correlation is present in the simulated data set.

Finally, one can clearly see an HBT-like enhancement for $q_{inv} < 50$~MeV/$c$ (solid green line), precisely in the region where it is expected. This structure, however, exhibits very similar behavior in both data sets, while it should be present only in case of real data (for the same reasons as $\pi^0-\pi^0$ correlation). Therefore, this structure can be attributed to artifacts of photon reconstruction procedures, which overshadows potential Bose-Einstein correlations. This can be further investigated by constructing the ratio between the correlation functions from real data and simulation, which should highlight any potential enhancement in the correlation region. Such a ratio for the left hand side of the $\pi^0$ peak is comparable to a constant value of $\sim 1.015$  (as shown in Fig.~\ref{fig:doubleRatio}) and exhibits no significant hint of a structure for $q_{inv} < 50$~MeV/$c$ that could be interpreted as Bose-Einstein correlation. Therefore, no significant direct photon signal can be extracted from the data. However, based on the estimates of statistical and systematic uncertainties, an upper limit can be determined.

\subsection{Upper limit on the $\lambda$ parameter}
\label{ssec:limit}
\begin{figure}
    \begin{minipage}{\columnwidth}
	        \centering
	        \includegraphics[width=\columnwidth]{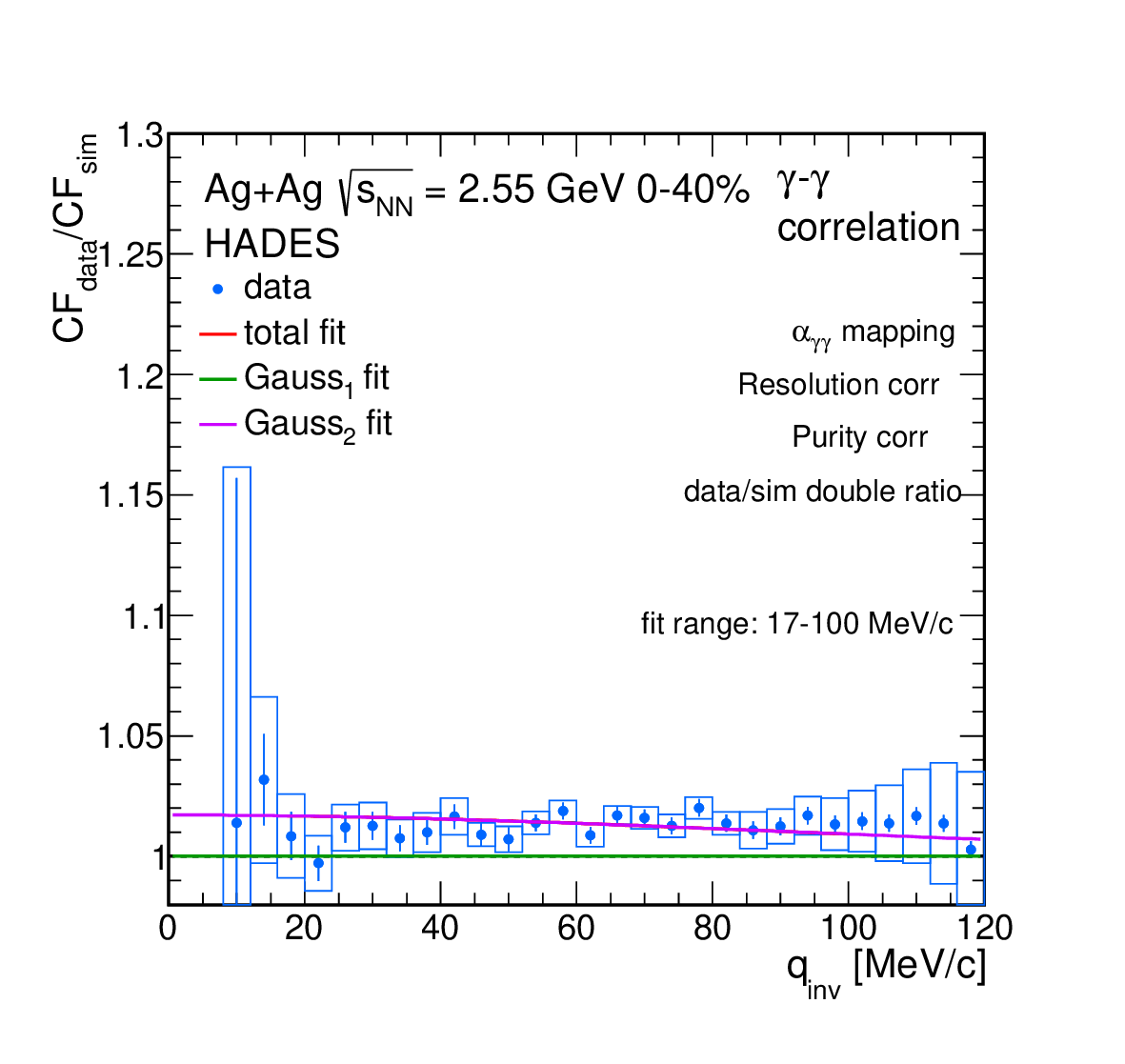}
	    \end{minipage}
    \caption{The ratio of the correlation functions from real and simulated data in $0 - 40$~\% most central Ag+Ag collisions at $\sqrt{s_{NN}} = 2.55$~GeV, fitted with the formula given in Eq.~\eqref{eq:doubleGauss}. Boxes around the data points represent systematic uncertainties. The purple line depicts the residual $\pi^0-\pi^0$ correlation contribution and the green line shows the expected direct photon Bose-Einstein correlation. The solid red line, fully covered by the purple one, represents the total double Gaussian function.}
    \label{fig:doubleRatio}
\end{figure}

The ratio of correlation functions from Fig.~\ref{fig:doubleRatio} was fitted with a double Gaussian function, defined as:
\begin{equation}
	CF_{ratio}(q_{inv}) = 1 + \lambda_{1}\,e^{-R_{1}^{2}\,q_{inv}^{2}} + \lambda_{2}\,e^{-R_{2}^{2}\,q_{inv}^{2}} 
	\label{eq:doubleGauss}
\end{equation}
where all of the parameters are the same as for Eq.~\eqref{eq:fitFunction}. 
Such a fit allows for an estimation of the $\lambda_{2}$ parameter without interference of low $q_{inv}$ detector-related enhancement (visible in the Fig.~\ref{fig:CFfinal}), hence allowing to quantify the residual correlation contribution more accurately. For fixed values of $R_{2}$ and $R_{1}$ this results in $\lambda_{1}\:\cdot\:10^{2} = 0 \pm 0.06$ (stat. uncert.) and $\lambda_{2}\:\cdot\:10^{2} = 1.71 \pm 0.12$ (stat. uncert.).

An upper limit on the $\lambda_{1}$ parameter can be determined for a given confidence level ($CL$) following a method developed by Feldman and Cousins \cite{FeldmanCousins}. The $R_{2}$ and $\lambda_{2}$ parameters are treated as fixed background constants with values obtained from the double Gaussian fit (see. Eq.~\eqref{eq:doubleGauss}). Two approaches for the treatment of $R_{1}$ were tested - treating it as a nuisance parameter with boundaries in the interval [2;7] fm or fixing it to a value obtained from a fit to real data (see Table~\ref{tab:parsExp}). The nuisance parameter approach yielded values of $R_{1}$ equal to either the upper or lower boundary, suggesting no true fit minimum was found.  Therefore, it was chosen to treat $R_{1}$ as a fixed parameter.

\begin{figure}
	\begin{minipage}{\columnwidth}
		\centering
		\includegraphics[width=\columnwidth]{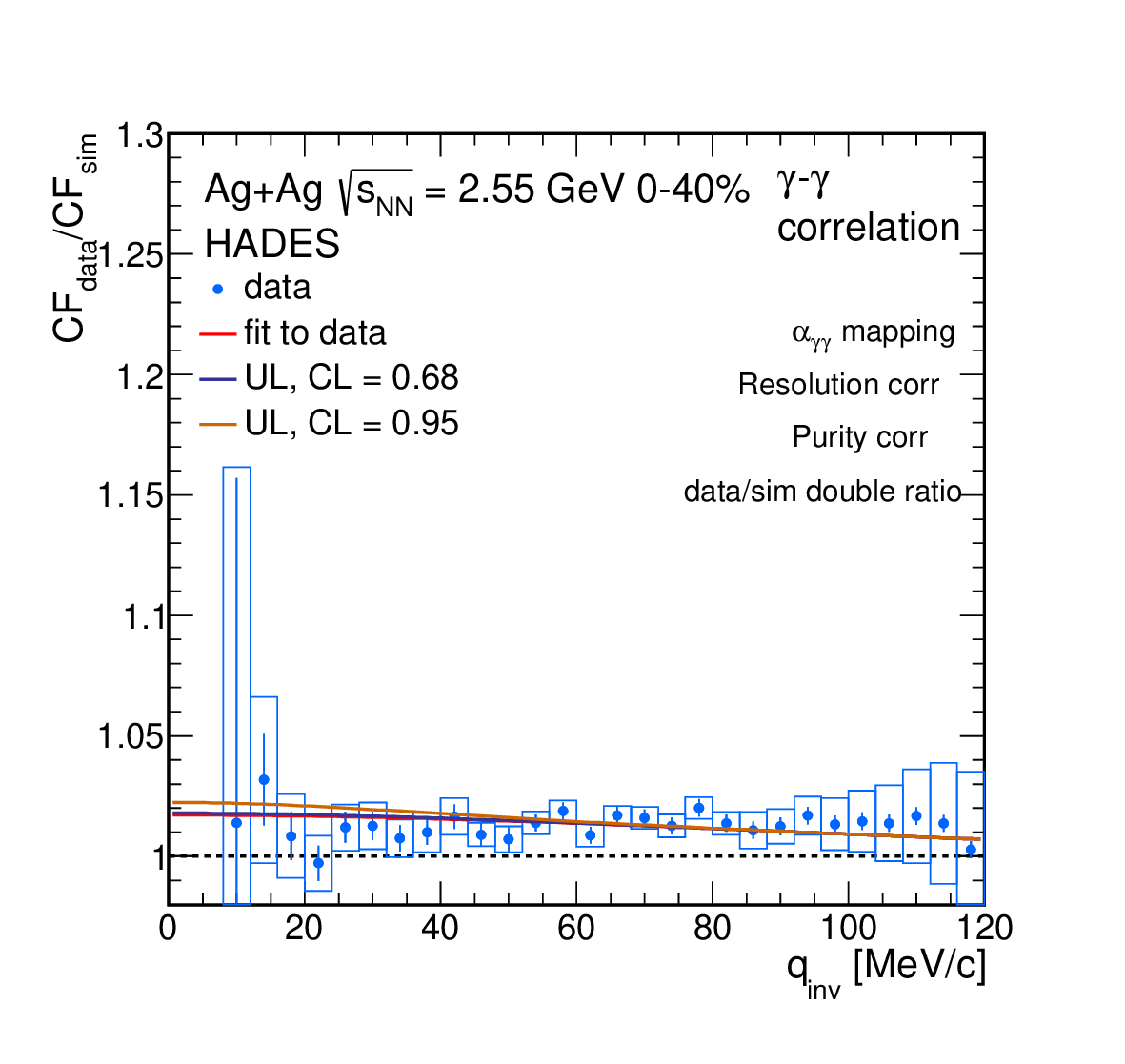}
	\end{minipage}
	\caption{The ratio of the correlation functions from real and simulated data in $0 - 40$~\% most central Ag+Ag collisions at $\sqrt{s_{NN}} = 2.55$~GeV. Boxes around the data points represent systematic uncertainties. All trend lines follow Eq. \eqref{eq:doubleGauss}. The red line represents fit to data (same as in the Fig. \ref{fig:doubleRatio}). The dark blue line corresponds to $\lambda_{1}$ value equal to upper limit with $CL = 0.68$, orange line  corresponds to $CL = 0.95$.}
	\label{fig:prediction}
\end{figure}

The numerical scan of plausible values of $\lambda_{1}$ was made in the range $[10^{-5};10^{-2}]$ with a step size of $10^{-6}$. For each $\lambda_{1}$ value $N=10^{5}$ ensemble correlation functions ($CF_{var}$) were generated by varying each point of the $CF_{pred}$ function, described by Eq.~\eqref{eq:doubleGauss}, by $\delta^{stat}$ and $\delta^{syst}$:

\begin{equation}
	\begin{split}
		CF_{pred} = CF_{ratio}(q_{inv},\lambda_{1})\\
		\delta^{stat}_{i} = \mathcal{N}(0,\,\sigma^{stat}_{i})\\  \delta^{syst}_{i} = \mathcal{N}(0,\,\sigma^{syst}_{i})\;
	\end{split}
	\label{eq:ULvariations}
\end{equation}

\begin{equation}
	CF_{var ~i} = CF_{pred ~i} + \delta^{stat}_{i} + \delta^{syst}_{i}
\end{equation}
where $i$ is an index to a given point of the function, $\mathcal{N}$ stands for the Gaussian distribution, $\sigma^{stat}$ is the statistical uncertainty and $\sigma^{syst}$ is the systematical uncertainty of a given point .

\begin{table}[h]
	\centering
	\caption{Upper limit predictions for $\lambda_{1}$ parameter, corresponding to graphical representation in the Fig.~\ref{fig:prediction}. See text for definitions and details.}
	\label{tab:parsUpperLimit}    
	\begin{tabular*}{\columnwidth}{@{\extracolsep{\fill}}lllll@{}}
		\hline
		parameter & fit to data & $UL, ~CL = 0.68$ & $UL, ~CL = 0.95$ \\
		\hline
		$\chi^2$ & $33.13$ & $34.47$ & $43.60$\\
		$\lambda_{1}\cdot10^{2}$ & $0 \pm 0.06$ & $0.08$ & $0.53$ \\
		\hline
	\end{tabular*}
\end{table}

Two confidence levels were employed: $CL = 0.68$ and  $CL = 0.95$, corresponding to 1$\sigma$ and 2$\sigma$ respectively. Table~\ref{tab:parsUpperLimit} shows the results for $\chi^2$ and $\lambda_{1}$ parameter for both $CL$ values, as well as for the fit presented in the Fig.~\ref{fig:doubleRatio}, serving as a reference. Fig. \ref{fig:prediction} shows hypothetical correlation functions in comparison to the data points from the Fig.~\ref{fig:doubleRatio}.

Based on Eq.~\eqref{eq:yield} and the results from Table~\ref{tab:parsUpperLimit} one can infer the $N_{\gamma, \; direct} / N_{\gamma, \; all}$ ratio, being below $0.04$ and $0.10$ for $CL = 0.68$ and  $CL = 0.95$ respectively.

\section{Summary}
\label{sec:summary}

Photon-photon correlations were measured in Ag+Ag collisions using the electromagnetic calorimeter of the HADES experiment. For the first time such a measurement was attempted in the region of a few GeV projectile kinetic energy. Experimental results were compared with the outcome of SMASH+HGeant simulations. Two-particle effects were estimated and suppressed to a point which allowed for an observation of a structure in the low $q_{inv}$ region, accompanied by an additional, weaker background-like contribution. The modified version of a Gaussian parametrization was used to fit the complex structure of the correlation function. The low $q_{inv}$ structure, despite similarity to the expected Bose-Einstein correlation, is most likely caused by the artifacts of particle reconstruction. An upper bound of a direct photon signal strength for two confidence levels was estimated as $\lambda_1 = 0.8 \cdot 10^{-3}$ at $CL = 0.68$ and $\lambda_1 = 0.53 \cdot 10^{-2}$ at $CL = 0.95$, along with the ratio of direct to all photons.

\begin{acknowledgements}
	\phantom{none}
	\newline
The collaboration gratefully acknowledges the support by

\begin{sloppypar}
SIP JUC Cracow, Cracow (Poland), 2017/26/M/ST2/00600, 2023/49/B/ST2/00652 (OPUS), SONATA-BIS 2023/50/E/ST2/00673; WUT Warsaw (Poland) No: 2020/38/E/ST2/00019 (NCN), IDUB-POB; IFD UW Warsaw (Poland) No: 2024/53/N/ST2/02107 (NCN); TU Darmstadt, Darmstadt (Germany), VH-NG-823, DFG GRK 2128, DFG CRC-TR 211, BMBF:05P18RDFC1, HFHF (Campus Darmstadt), ELEMENTS 500/10.006, GSI F\&E, EMMI GSI Darmstadt; Goethe-University, Frankfurt (Germany), BMBF:05P12RFGHJ, GSI F\&E, HFHF (Campus Frankfurt), EMMI GSI Darmstadt, ELEMENTS 500/10.006; BU-Wuppertal, Wuppertal (Germany), BMBF 05P24PX1; JLU Giessen, Giessen (Germany), BMFTR 05P24RG6; IJCLab Orsay, Orsay (France), CNRS/IN2P3; NPI CAS, Rez, Rez (Czech Republic), MSMT LM2023060, MSMT OP JAK Z.02.01.01/00/23$_{-}$015/0008181; Frederick Research Center, Nicosia (Cyprus), EXCELLENCE HUBS: EXCELLENCE/0925/227, under THALIA 2021-2027, co-funded by the EU through RIF;
This experiment was part of the FAIR-Phase0 program. The publication is funded by the OpenAccess Publishing Fund of GSI Helmholtzzentrum fuer Schwerionenforschung.
We acknowledge the contribution from the following colleagues:: Alexander Belyaev, Oleg Fateev, Alexander Ierusalimov, Vladimir Ladygin, Sergei Reznikov, Alexander Zinchenko, Dmitriy Borisenko, Marina Golubeva, Fedor Guber, Alexander Ivashkin, Nikolay Karpushkin, Sergey Morozov, Oleg Petukhov, Andrei Reshetin, Arseniy Shabanov, Vladimir Khomyakov, Alexander Lebedev, Alexander Zhilin, Oleg Golosov, Mikhail Mamaev, Arkadiy Taranenko.
\end{sloppypar}

\end{acknowledgements}

\bibliographystyle{spphys}    
\bibliography{bibliography}   

\end{document}